\documentclass{article}\usepackage[]{graphicx}\usepackage[]{color}
\makeatletter
\def\maxwidth{ %
  \ifdim\Gin@nat@width>\linewidth
    \linewidth
  \else
    \Gin@nat@width
  \fi
}
\makeatother

\usepackage{framed}
\makeatletter
 {\par\unskip\endMakeFramed%
 \at@end@of@kframe}
\makeatother

\usepackage{alltt} 

\usepackage{etoolbox}
\newtoggle{arxivformat}
\toggletrue{arxivformat}

\iftoggle{arxivformat} {
  \usepackage[sort&compress, numbers]{natbib}
  \usepackage{times}
}

\usepackage{hyperref}

\usepackage[margin=1.5in]{geometry}
\usepackage{url}
\usepackage{xcolor}
\usepackage{colortbl}
\usepackage{fancybox}
\usepackage{hyperref}
\usepackage{graphicx} 
\usepackage{subfigure}
\usepackage{bbm}
\usepackage{caption}
\definecolor{newcolor}{rgb}{.8,.349,.1}

\title{Topic representation: finding more representative words in topic models}

\usepackage[scaled=0.7]{beramono}

\author{
Jinjin Chi\\
Jilin University\\
\and
Jihong Ouyang \\
Jilin University\\
\and
Changchun Li \\
Jilin University\\
\and
Xueyang Dong\\
Jilin University\\
\and
Ximing Li\\
Jilin University\\
\texttt{liximing86@gmail.com}
\and
Xinhua Wang\\
Jilin University\\
}

\iftoggle{arxivformat} {
}{
  \nipsfinalcopy 
}
\IfFileExists{upquote.sty}{\usepackage{upquote}}{}
\begin{document}

\maketitle

\begin{abstract}
The top word list, i.e., the top-\emph{M} words with highest marginal probability in a given topic, is the standard topic representation in topic models. Most of recent automatical topic labeling algorithms and popular topic quality metrics are based on it. However, we find, empirically, words in this type of top word list are not always representative. The objective of this paper is to find more representative top word lists for topics. To achieve this, we rerank the words in a given topic by further considering marginal probability on words over every other topic. The reranking list of top-\emph{M} words is used to be a novel topic representation for topic models. We investigate three reranking methodologies, using (1) standard deviation weight, (2) standard deviation weight with topic size and (3) Chi Square $\chi ^2$ statistic selection. Experimental results on real world collections indicate that our representations can extract more representative words for topics, agreeing with human judgements.

\end{abstract}


\section{Introduction}\label{sec:intro}

Probabilistic topic modeling family \citep{Blei2012} has become a mainstream tool for analyzing the text document collection. This model family assumes that each document is a mixture of latent topics, where each topic is a multinomial distribution over the vocabulary.

Topic models such as latent Dirichlet allocation \citep{Blei2003} have empirically achieved great success in modeling documents so far. With algorithms for approximate posterior inference, we can use topic models to uncover latent variables with respect to topics from a collection of documents, leading to semantically meaningful decompositions of them. Topics place high probability on words to represent concepts, and documents are described by mixtures of these concepts. Due to the success in discovering semantics knowledge, topic models are usually used in natural language processing tasks, such as multi-document summarisation \citep{Sum2009} and novel word sense detection \citep{WordSense2012}.

Perusing the learnt topics is the core mission in topic modeling for documents. The standard topic representation is the top word list, i.e., the top-\emph{M} words with highest marginal probability in a given topic. To simplify the topic representation, some attempts aim at automatically labeling topics, generating labels that can explicitly identify the semantics of topics. For example, the first attempt, to our knowledge, is proposed in \citep{Labeling2007}, which generates candidate labels from a reference collection using noun chunks and bigrams with high lexical association. The authors of \citep{labeling2011} suggest an automatical topic labeling algorithm using Wikipedia article titles to process candidate labels. More recently, an interesting algorithm \citep{Labeling2013} represents topics by image labels, instead of text labels. PageRank is used to select the most suitable candidate images.

Among these existing topic representations, the top word list is acknowledged to be the basic representation for topics, and most of alternatives, to our knowledge, are based on it. More broadly, the automatical topic quality evaluation metrics \citep{Intrusion2009,Coh2011,Eva2013} are also based on the top word list. For example, the popular topic coherence metric is computed by counting the co-occurrence numbers among top words in a given topic, following the intuition that more frequently co-occurring intends more coherent a topic is.

All algorithms based on the top word list mentioned above, i.e., both alternative topic representations and topic quality metrics, follow a basic assumption that the top-\emph{M} words ranked by the topic-word distributions are the most representative words for topics. However, that is not always the case. With statistics inference algorithms, the words with highest marginal probability should be high frequency words, no matter whether these words are on a specific subject. At worst, a very frequently occurring but meaningless word will be the top word for most of topics, just like the stopwords. Some examples can be seen in the next section. Unfortunately, due to the power-law characteristics of language, it is impossible to make a clean sweep of such stopword-like words in practice.

To address the problem mentioned above, we aim to find more representative words for topics, and use the novel list of top-\emph{M} words to represent topics. This objective can be achieved by reranking the words in a given topic by further considering marginal probability on words over every other topic. Considering two cases: (1) if a top word in a given topic is also in the top word lists for most of other topics, this word is a bad top word, which is in fact unrepresentative; (2) if a word is not occurring very frequently but most of occurrences are assigned to a same topic, this word might be representative for that topic. Following the analysis, we propose three reranking methodologies, using (1) standard deviation weight, (2) standard deviation weight with topic size and (3) Chi Square $\chi ^2$ statistic selection. We conduct a number of experiments to evaluate our algorithms on real world collections. Empirical results indicate that in contrast to the standard top word list with highest marginal probability, our algorithms can extract more representative words for topics, agreeing with human judgements.

\section{Problem Description} \label{sec:lr}
In this section, we present the problems in the standard top word list representation. We first review some preliminaries, including two topic models and a popular topic quality metric. The studied topic models are latent Dirichlet allocation (LDA) \citep{Blei2003} and Multi-grain clustering topic model (MGCTM) \citep{MGCTM2013}; and the topic quality metric is the so called topic coherence \citep{Coh2011}.
\subsection{Preliminaries}
\paragraph{LDA} LDA is a generative probabilistic model for the text document collection. This model consists of \emph{K} topics, where each topic is a multinomial distribution $\phi$ over the vocabulary, drawn from a Dirichlet prior $\beta$. To generate a document \emph{d}, LDA first draws a topic mixture proportion $\theta_d$ from a Dirichlet prior $\alpha$, and then repeatedly generates word tokens by sampling a topic indicator ${z_{dn}}$ from the distribution $\theta _d$ and then sampling a word $w _{dn}$ from the selected topic distribution $\phi _{z _{dn}}$.

\paragraph{MGCTM} MGCTM extends LDA by dividing topics into two categories. One is the global topic used to capture corpus-level common semantics; the other is the local topic used to capture document-level specific semantics. Under MGCTM, there exist \emph{R} global topics, and local topics are organized into \emph{J} \emph{K}-sized latent groups. To generate a document \emph{d}, (1) draw a multinomial distribution $\theta ^g_d$ over global topics, from the global Dirichlet prior $\alpha ^g$; (2) choose a group $\eta_d$ and draw a multinomial distribution $\theta ^l_d$ over local topics, from the selected group-specific Dirichlet prior $\alpha ^l_{\eta_d}$; (3) draw a Bernoulli decision distribution $\omega_d$ from the Dirichlet prior $\gamma$. To generate a word token, first sample a binary variable $\delta_{dn}$ from the Bernoulli decision distribution $\omega_d$; If $\delta_{dn}=1$, this word token $w_{dn}$ will be generated from the global topic mixture proportion $\theta ^g_d$, otherwise $w_{dn}$ will be generated from the local topic mixture proportion $\theta ^l_d$. The subsequent word generative process is the same as LDA. Profiting from the global topic design, MGCTM can uncover the common semantics and filter out the noise words (e.g., stopword-like words) in some degree.

\paragraph{Topic coherence} Topic coherence is a very popular automatical metric to evaluate whether the topics learnt by topic models are coherent. The intuition behind this metric is that a topic is more coherent if its most representative words are more frequently co-occurring. Given the top-\emph{M} word list ${V^k} = \left( {v_1^k, \cdots ,v_M^k} \right)$ in the topic distribution $\phi_k$, the coherence value of this topic \emph{k} is computed by:
\begin{equation}\label{Eq1}
Coh(k,V^k) = \sum\limits_{m = 2}^M {\sum\limits_{l = 1}^{m - 1} {\log \frac{{D\left( {v_m^k,v_l^k} \right) + \varepsilon }}{{D\left( {v_l^k} \right)}}} }
\end{equation}
where $D\left( v \right)$ is the number of documents containing the word type \emph{v}; $D\left( {{v_1},{v_2}} \right)$ is the number of documents containing both word type $v_1$ and $v_2$; $\varepsilon$ is a smoothing constant used to avoid log zero. For topic coherence, values closer to zero imply greater co-occurrence, i.e., better coherent performance.

\subsection{Problem}

The story begins with an experiment on Twitter, where our original intention is to investigate topic modeling on short texts. Short texts contain very a few word tokens and are commonly quite noisy. Documents from Twitter are typical short texts. We collect 1,000,000 Twitter documents from the web\footnote{After standard per-processing such as removal of stopwords, we obtain a Twitter collection with a 95,623-sized vocabulary.}; and then fit LDA and MGCTM\footnote{For LDA, the number of topics is set to 50; for MGCTM, the number of latent groups is set to 25, where each group contains 2 local topics, and the number of global topics is set to 5.} on this Twitter collection respectively in order to evaluate whether MGCTM can filter out the noise words in short texts. Surprisingly, we observe a very interesting result. Table 1 presents two top-10 highest probability words of corresponding topics learnt by LDA and MGCTM respectively. Their topic coherence values are listed in the first column. We observe that in contrast to LDA, MGCTM can effectively filter out some noise words (e.g., ``lol'' and ``rt'' in box\footnote{The word ``lol'' is the abbreviation of ``laugh out loud'', used as a modal particle; and the word ``rt'' is the abbreviation of ``retweet''. Because they are too common in Twitter and lack specific semantics, we consider them as noise.}), but its topic coherence values are even worse than LDA's (e.g., -256.5 vs. -255.7). This leads to an obvious conflict between human knowledge and the topic coherence metric.

Reviewing the topic word lists learnt by LDA and MGCTM in Table 1, we find that the only difference between corresponding topics is the two noise words ``lol'' and ``rt''. Statistics results\footnote{The average document length of \emph{Twitter} is only 4.2, however, ``lol'' occurs in 100K/1000K texts and ``rt'' occurs in 80K/1000K texts.} show that both of them are much more frequently occurring than other top words, so they are more co-occurring with other top words in general, resulting in better topic coherence values. Based on this analysis, we believe that the conflict in topic coherence is caused by this kind of frequently occurring but meaningless words (i.e., just like the stopwords), existing in the top word list.

To further support our analysis, we conduct an additional experiment on the \emph{Newsgroup} collection (See more detail about this collection in Section 4). We prepare two versions of \emph{Newsgroup}, where one is the original collection (abbr. S-Ng) and the other is a processed collection (abbr. NS-Ng) by removing the stopwords. We simultaneously fit 50-topic LDA models on S-Ng and NS-Ng, and present two top-10 highest probability words of corresponding topics in Table 2. Unsurprisingly, we still see the conflict in the topic coherence metric mentioned above.

\definecolor{mymygray}{rgb}{0.97,0.97,0.97}
\definecolor{mygray}{rgb}{0.92,0.92,0.92}
\begin{table}[t]
\renewcommand\arraystretch{1.5}
\caption{Two top-10 word lists of corresponding topics learnt by LDA and MGCTM, respectively. The first and third rows are topics learnt by LDA; the second and fourth rows are topics learnt by MGCTM. The first column shows topic coherence values.}
\label{sample-table}
\vskip 0.15in
\begin{center}
\begin{small}
\begin{tabular}{p{0.2\columnwidth}p{0.65\columnwidth}}
\hline
Topic coherence & Top-10 word list\\
\hline
\rowcolor{mymygray}
-255.7    &   good, night, day, morning, sleep, time, work, \ovalbox{lol}, today, home \\
\rowcolor{mygray}
-256.5    &	good, day, morning, night, today, time, work, tomorrow, sleep, home \\\hline
\rowcolor{mymygray}
-282.6	&  iphone,	ipad, apple, free, app, \ovalbox{rt},	android, phone,	online,	buy		 \\
\rowcolor{mygray}
-297.4	&  iphone,	ipad, apple, phone, follow, free, app, android, buy, online	 \\
\hline
\end{tabular}
\end{small}
\end{center}
\vskip -0.1in
\end{table}

\begin{table}[t]
\renewcommand\arraystretch{1.5}
\caption{Two top-10 word lists of corresponding topics learnt by LDA on S-Ng and NS-Ng, respectively. The first and third rows are topics learnt on S-Ng; the second and fourth rows are topics learnt on NS-Ng. The first column shows topic coherence values.}
\label{sample-table}
\vskip 0.15in
\begin{center}
\begin{small}
\begin{tabular}{p{0.2\columnwidth}p{0.65\columnwidth}}
\hline
Topic coherence & Top-10 word list\\
\hline
\rowcolor{mymygray}
-66.3   & the, of, space, you, are, to, on, for, nasa, and\\
\rowcolor{mygray}
-179.6  & space, nasa, earth, launch, gov, orbit, moon, shuttle, satellite, mission \\\hline
\rowcolor{mymygray}
-89.8   & of, to, and, a, is, medical, for, disease, with, in		 \\
\rowcolor{mygray}
-188.6  & health, medical, insurance, disease, doctor, treatment, patients, care, medicine, drug \\
\hline
\end{tabular}
\end{small}
\end{center}
\vskip -0.1in
\end{table}

\section{Methodology}

Before introducing methodologies, we first propose basic settings and symbol definitions. In this work, we investigate novel topic representations around LDA and use collapsed Gibbs sampling (CGS) \citep{Gibbs2004} for approximate LDA posterior inference. CGS involves sequentially resampling each topic assignment $z_{dn}$ from its conditional posterior, holding all other variables fixed. Given final samples of $z_{dn}$, the point estimates of the topic-word distributions $\phi$ can be computed by:
\begin{equation}\label{Eq2}
\phi _{kv} = \frac{{N_{kv} + \beta }}{{N_k + V\beta}}
\end{equation}
where $N_{kv}$ and $N_k$ are the number of the word type \emph{v} assigned to the topic \emph{k} and the total number of word tokens assigned to the topic \emph{k}, respectively; \emph{V} is the number of word types. Moreover, Let $N_v$ be the number of the word type \emph{v} has occurred; $N$ be the total number of word tokens has occurred in a collection.

In terms of Eq.2, given a topic \emph{k}, ranking top-\emph{M} words by the topic distribution $\phi _k$ is equivalent to finding \emph{M} words with the largest $N_{kv}$ values. This mechanism favors high frequency words, i.e., words with large $N_v$ values. Frequently occurring but meaningless words, e.g., stopwords, sometimes mingle in the top-\emph{M} word list, resulting in bad topic representation.

We consider that for a given topic \emph{k}, a representative word \emph{v} should be not only (1) with high marginal probability (i.e, large $\phi_{kv}$ values), but also (2) with low marginal probabilities in every other topic (i.e., small $\phi_{tv}$ values, where $t \in \left\{1,2,\cdots, K \right\} _{\neg k}$). The standard top-\emph{M} word list representation, ranked by the topic distributions $\phi$, neglects the second factor. To improve it, we further consider the second factor to rerank topic words for more representative words. We propose three reranking methodologies in this paper, using (1) standard deviation weight (SDW), (2) standard deviation weight with topic size (SDWTS) and (3) Chi Square $\chi ^2$ statistic selection (CHI).

\paragraph{SDW reranking} Given a topic \emph{k}, we provide weighting values for its distribution $\phi_k$ over all \emph{V} words by:
 \begin{equation}\label{Eq?}
weight_{SDW}(k,v) = \sqrt {{{\sum\limits_{i \ne k} {\left( {{\phi _{kv}} - {\phi _{iv}}} \right)} }^2}}
\end{equation}
This weight is a pseudo standard deviation. It treats the processed partner $\left( k,v \right)$, i.e., $\phi _{kv}$, as the expectation, and computes its standard deviation over the probabilities of the same word \emph{v} in every other topic. The intuition behind is that if a word \emph{v} with larger $\phi _{kv}$ value in the topic \emph{k} also corresponds to larger $\phi_{iv}$ values in most of other topics (i.e., $i\ne k$), we will provide a small weight to $\phi _{kv}$; otherwise, we will provide a large weight to $\phi _{kv}$. Finally, we can use the following weighted $\phi _{kv}$ values to rerank words in topics:
\begin{equation}\label{Eq?}
\phi _{kv}^{SDW} = weight_{SDW}(k,v) \times {\phi _{kv}}
\end{equation}

\paragraph{SDWTS reranking} Based on SDW reranking, we further consider the topic size $N_k$ (i.e., the number of word tokens assigned to each topic by Gibbs sampling). The topic size itself is a reasonable predictor of topic quality \citep{Coh2011}, where larger topic size implies better topic quality. Considering this, we present a novel weighting equation as follows:
\begin{eqnarray}\label{Eq?}
\!\!\!\!\!\!\!\!\!\!\!\!\!\!\!\!\!\!\!\!\!\!\!\!\!\!\! & &  weight_{SDWTS}(k,v) = \sqrt {{{\sum\limits_{i \ne k} {\left( {{\phi _{kv}}{N_k} - {\phi _{iv}}}{N_i} \right)} }^2}} \nonumber \\
\!\!\!\!\!\!\!\!\!\!\!\!\!\!\!\!\!\!\!\!\!\!\!\!\!\!\! & &   \quad \quad \quad \quad \quad \quad \quad \quad \quad \approx  \sqrt {{{\sum\limits_{i \ne k} {\left( N_{kv} - N_{iv} \right)} }^2}}
\end{eqnarray}
Finally, we can use the following weighted $\phi _{kv}$ values to rerank words in topics:
\begin{equation}\label{Eq?}
\phi _{kv}^{SDWTS} = weight_{SDWTS}(k,v) \times {\phi _{kv}}
\end{equation}
Reviewing Eq.5, we note that the SDWTS weight is focusing on the number of word tokens assigned to each topic directly.

\paragraph{CHI reranking} Chi Square $\chi ^2$ statistic has been widely used in feature selection for classification. It selects most discriminative features by measuring the statistical dependency between the feature and the category. In our case, we can treat words/topics as features/categories, and then directly use Chi Square $\chi ^2$ statistic to rank words in topics. The $\chi ^2$ with \emph{V} different words and \emph{K} topics is defined as:
\begin{equation}\label{Eq?}
{\chi ^2} = \sum\limits_{k = 1}^K {\sum\limits_{v = 1}^V {\frac{{{{\left( {{N_{kv}} - {E_{kv}}} \right)}^2}}}{{{E_{kv}}}}} }
\end{equation}
where ${E_{kv}} = {{{N_k}{N_v}} \mathord{\left/{\vphantom {{{N_k}{N_v}} N}} \right. \kern-\nulldelimiterspace} N}$. The $\chi ^2$ value for the word \emph{v} in the topic \emph{k} can be interpreted by the following probability:
\begin{equation}\label{Eq?}
{\chi ^2}\left( {k,v} \right) = \frac{{N{{\left( {p\left( {k,v} \right)p\left( {\neg k,\neg v} \right) - p\left( {\neg k,v} \right)p\left( {k,\neg v} \right)} \right)}^2}}}{{p\left( k \right)p\left( {\neg k} \right)p\left( v \right)p\left( {\neg v} \right)}}
\end{equation}
where $p\left( {k,v} \right)$ is the probability of the topic \emph{k} containing the word \emph{v} and $p\left( {\neg k,\neg v} \right)$ is the probability of not being in the topic \emph{k} and not containing the word \emph{v} and so on. Given all $\chi ^2$ values, we can rerank the vocabulary for every topic.

\section{Evaluation}

In this section, we evaluate our novel topic representations on two real world collections. The first is \emph{Newsgroup}\footnote{http://web.ist.utl.pt/$~$acardoso/datasets/}, a collection of news releases, which contains 18,821 documents with a 93,864-sized vocabulary. We generate two versions of \emph{Newsgroup}, where one is the original collection (abbr. S-Ng) and the other is a processed collection (abbr. NS-Ng) by removing the stopwords. The second collection is \emph{Wikipedia}\footnote{http://topics.cs.princeton.edu/nubbi/nubbi\_data\_censored.tar.gz} (abbr. Wiki), which contains 1,918 documents. After pre-processing, i.e., word stemming and removal of the stopwords, 9,144 word types are left.

In our experiments, the top-\emph{M} word list ranked by $\phi$ is called as NORM representation, and the top-\emph{M} word list reranked by SDW/SDWTS/CHI methodology is called as SDW/SDWTS/CHI representation.

\subsection{Evaluation on Stopwords Filtering}

The first evaluation is on whether our reranking representations can filter out stopword-like words in the top list (i.e., top words occurring in most of topics). For this goal, we fit a 100-topic LDA model on NS-Ng, and randomly select two topics, which are about food and baseball game, for visualization. The top-20 word lists of all four topic representations are illustrated in Table 3.

The first row is the top word list about food. Overall, we see that all four representations are in the main coherent. The word ``food'' is always the best representative word. However, in the NORM representation, some exalted words, such as ``nasa'' and ``writes'', are less related to the topic about food, and moreover, some words, such as ``don'' and ``articles'', seem the stopword-like words. The word ``don'' is obviously stopword-like. It is the first half of the stopword ``don't'', but is lucky to be left due to the misrecognition of the right single quote. We are a little surprised to judge that the word ``articles'' is stopword-like, because it has occurred in 35 top lists out of 100 topics, much more frequent than most of other top words. Compared to the NORM representation, the top word lists of the three reranking representations seem better.

The second row is the top word list about baseball game. In the NORM representation, again we see that the top words include some less relevant words, such as ``ride'' and ``technology'', and the two stopword-like words, i.e., ``don'' and ``articles''. Besides, its top one word ``year'' is less representative for this topic. In contrast, the three reranking representations seem better. First, they all successfully filter out the two stopword-like words. Second, they are obviously more coherent about the subject baseball game.

\begin{table}[t]
\renewcommand\arraystretch{1.8}
\caption{Top-20 word lists learnt by LDA on NS-Ng. The first row is the topic about ``food'' and the second row is the topic about ``baseball game''.}
\label{sample-table}
\vskip 0.15in
\begin{center}
\begin{small}
\begin{tabular}{p{0.21\columnwidth}|p{0.21\columnwidth}|p{0.21\columnwidth}|p{0.21\columnwidth}}
\hline
NORM & SDW & SDWTS & CHI\\
\hline
food, gov, nasa, writes, eat, article, foods, don, apr, fat, chinese, sensitivity, brain, taste, jpl, eating, people, reaction, related, superstition & food, foods, eat, sensitivity, chinese, fat, taste, gov, restaurant, vegetables, nasa, eating, glutamate, dealy, meat, larc, milk, gsfc, allergic, brain
& food, foods, eat, sensitivity, chinese, fat, taste, gov, restaurant, vegetables, nasa, eating, glutamate, dealy, meat, larc, milk, gsfc, allergic, elroy
& food, foods, eat, chinese, eating, sensitivity, fat, taste, gov, restaurant, glutamate, dealy, meat, milk, allergic, brain, seizure, sugar, vegetables. diet\\\hline
year, game, baseball, games, players, team, articles, runs, season, ride, technology, league, player, average, don, play, apr, pitcher, pitching, sox
& baseball, game, players, braves, pitcher, pitching, sox, year, cubs, runs, games, jays, mets, season, team, phillies, pitchers, morris, alomar, pitch
& baseball, game, players, braves, pitcher, pitching, innings, year, cubs, runs, games, jays, mets, season, team, phillies, pitchers, sox, alomar, pitch
& baseball, game, players, braves, pitcher, pitching, sox, cubs, runs, year, games, jays, mets, season, phillies, pitchers, morris, alomar, pitch, team \\\hline
\end{tabular}
\end{small}
\end{center}
\vskip -0.1in
\end{table}

We are further interested in whether our reranking representations can filter out the true stopwords in the top word list. For this purpose, we fit a 100-topic LDA model on S-Ng, and still visualize the two topics about food and baseball game. The top-20 word lists of all four topic representations are illustrated in Table 4.

We can observe that in the NORM representation, the top words are almost the stopwords, e.g., the topic about food contains 18/20 stopwords and the topic about baseball game contains 13/20 stopwords. Such kind of topics must be useless. In contrast, our reranking representations perform significantly better. They successfully filter out even the true stopwords, which are much frequently occurring in every document. Besides, we see that all three reranking representations seem coherent. Only a few less relevant words exist in the range from the 10th to 20th top words.

\begin{table}[t]
\renewcommand\arraystretch{1.8}
\caption{Top-20 word lists learnt by LDA on S-Ng. The first row is the topic about ``food'' and the second row is the topic about ``baseball game''.}
\label{sample-table}
\vskip 0.15in
\begin{center}
\begin{small}
\begin{tabular}{p{0.21\columnwidth}|p{0.21\columnwidth}|p{0.21\columnwidth}|p{0.21\columnwidth}}
\hline
NORM & SDW & SDWTS & CHI\\
\hline
the, and, to, i, that, of, it, a, is, in, t, msg, you, have, food, these, or, this, not, are
& food, chinese, sensitivity, superstition, foods, taste, spdcc, restaurant, dyer, glutamate, eat, meat, cousineau, nmm, questor, compdyn, allergic, dougb, olney, sugar,
& food, chinese, sensitivity, superstition, foods, taste, carcinogenic, restaurant, dyer, glutamate, eat, meat, cousineau, nmm, questor, compdyn, allergic, dougb, olney, blah,
& food, chinese, sensitivity, superstition, foods, taste, spdcc, restaurant, dyer, glutamate, eat, meat, cousineau, nmm, allergic, questor, compdyn, dougb, olney, carcinogenic\\\hline
in, a, i, s, to, edu, is, that, they, year, game, baseball, but, hit, games, writes, this, will, be, and
& baseball, braves, hit, sox, magnus, gant, games, ohio, game, yankees, acs, reds, hitting, clutch, players, year, mets, dodgers, catcher, phillies
& baseball, braves, hit, sox, magnus, gant, games, ohio, game, yankees, acs, reds, hitting, clutch, players, year, mets, dodgers, catcher, phillies
& baseball, braves, hit, sox, magnus, gant, games, ohio, game, yankees, reds, acs, hitting, clutch, players, year, mets, dodgers, catcher, fans \\\hline
\end{tabular}
\end{small}
\end{center}
\vskip -0.1in
\end{table}

 Figure 1 lists the top-10 words in the topic about food. Because the stopwords are with higher $\phi$ values, they dominate the top-10 list. The more representative words are lower ranked, e.g., the order of the word ``foods'' is only 109. The SDWTS reranking successfully filters out the stopwords from the top-10 list by providing larger weights to the dedicated words. We can observe that the $\phi ^{SDWTS}$ values of the stopwords are much smaller than those of the final top words in the SDWTS representation.

\begin{figure}[t]
\vskip 0.2in
\begin{center}
\centerline{\includegraphics[width=0.6\columnwidth]{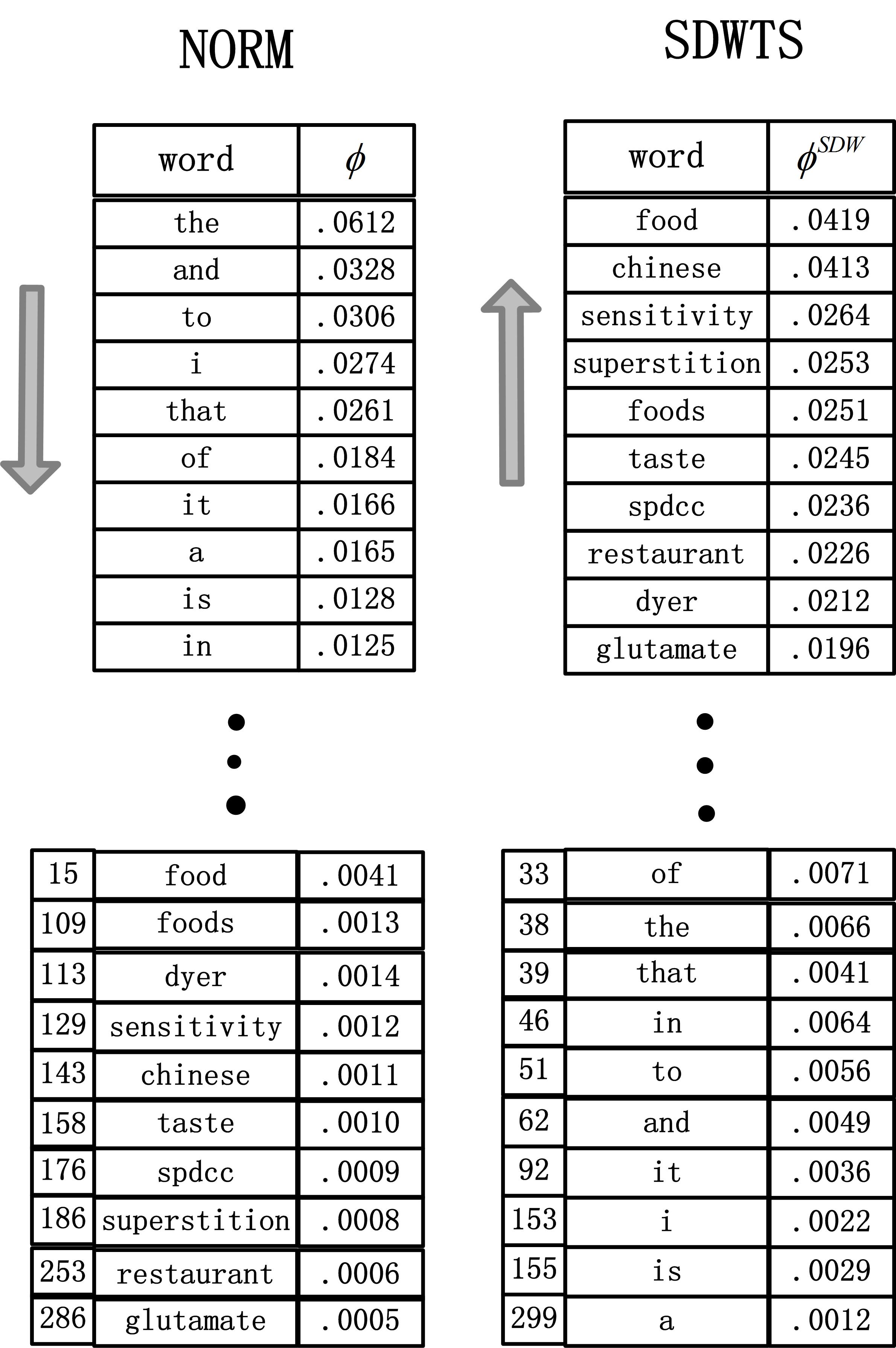}}
\caption{The reranking processing of the SDWTS representation for the topic about food. The first/four column in the bottom is the word order in the NORM/SDWTS representation.}
\label{Fig1}
\end{center}
\vskip -0.2in
\end{figure}

\subsection{Evaluation on Human-Interpretability}
\label{sec:4.2}

The second evaluation is on whether the top words in our reranking representations are more representative than in the NORM representation. Because there are no gold-standard top word lists of topics, we use the word intrusion task \citep{Intrusion2009} for indirect evaluation. In this task, each topic is presented by the top six words. Randomly remove one of six top words and then randomly select an intruder word to replace the removed top word. The task of users are asked to identify the intruder word. We consider that if the top topic words are more representative, it should be easier for users to identify the intruder word. Following this, for a same topic, more representative top words must lead to higher intrusion accuracy given by:
\begin{equation}\label{Eq1}
AC = \frac{{\sum\nolimits_{k = 1}^K {\mathbbm{1}( {{i_k} = {w_k}})} }}{K}
\end{equation}
where $i_k$ is the intruder word of the topic \emph{k} selected by a human being, and $w_k$ is the true intruder word of the topic \emph{k}. In our experiments, we use an automatical intruder detector \citep{AutoEva2014}, which can emulate the performance of human judgements. Three different patterns are used to generate the intruder words: selecting (1) from the vocabulary (S\_VOC), (2) from the top six words of other topics (S\_TOPIC) and (3) from the 11th to 100th words of the current topic (S\_SELF).

We fit a 150-topic LDA model on NS-Ng and a 100-topic LDA model on Wiki. See the following discussions on results of S\_VOC, S\_TOPIC and S\_SELF tests.

\paragraph{Performance on S\_VOC and S\_TOPIC tests} For both tests, we randomly generate the intrusion topics 10 times, and finally present the average intrusion accuracy.

The performance on S\_VOC test is shown in Figure 2. For NS-Ng, the NORM representation accuracy is about 0.8, and the three reranking representations are all above 0.9. For Wiki, accuracy values of all the four are over 0.93, and reranking representations perform better, i.e., about 0.97. Among the three, the SDWTS and CHI representations perform better than the SDW representation and more stable.

The performance on S\_TOPIC test is shown in Figure 3. Overall, we can see that the three reranking representations perform significantly better than the NORM representation. For NS-Ng, the accuracy values of SDW and CHI representations are about 0.85, the SDWTS accuracy is about 0.83, but the NORM accuracy is just about 0.77. For Wiki, The performance gap is even larger, e.g., the accuracy values of the SDW and SDWTS representations are about 0.9, but the NORM accuracy is even lower than 0.7. Among the three reranking representations, the CHI accuracy values are the best. We see that the performance of S\_TOPIC test is worse than that of S\_VOC, especially for the NORM representation. Table 5 shows two examples that the NORM representation loses.

\begin{figure}[t]
\vskip 0.2in
\begin{center}
\centerline{\includegraphics[width=1.0\columnwidth]{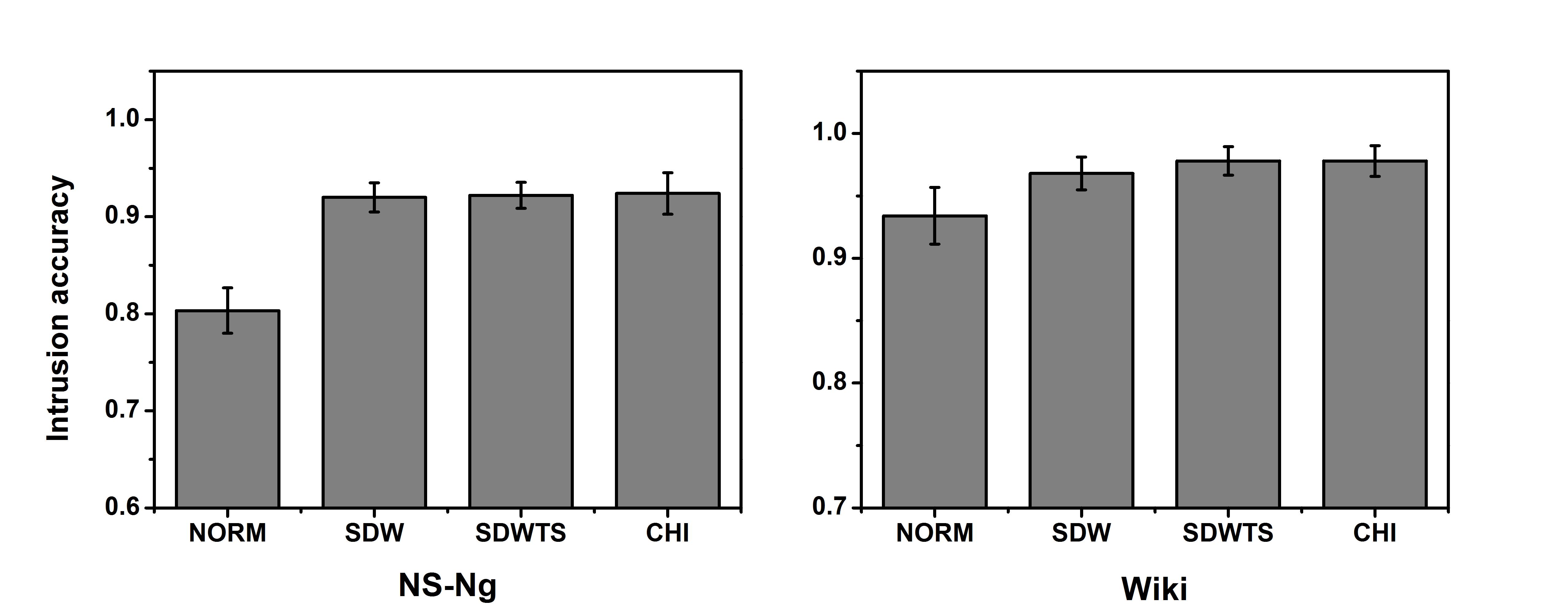}}
\caption{Performance of S\_VOC test on NS-Ng (left) and Wiki (right).}
\label{Fig1}
\end{center}
\vskip -0.2in
\end{figure}

\begin{figure}[t]
\vskip 0.2in
\begin{center}
\centerline{\includegraphics[width=1.0\columnwidth]{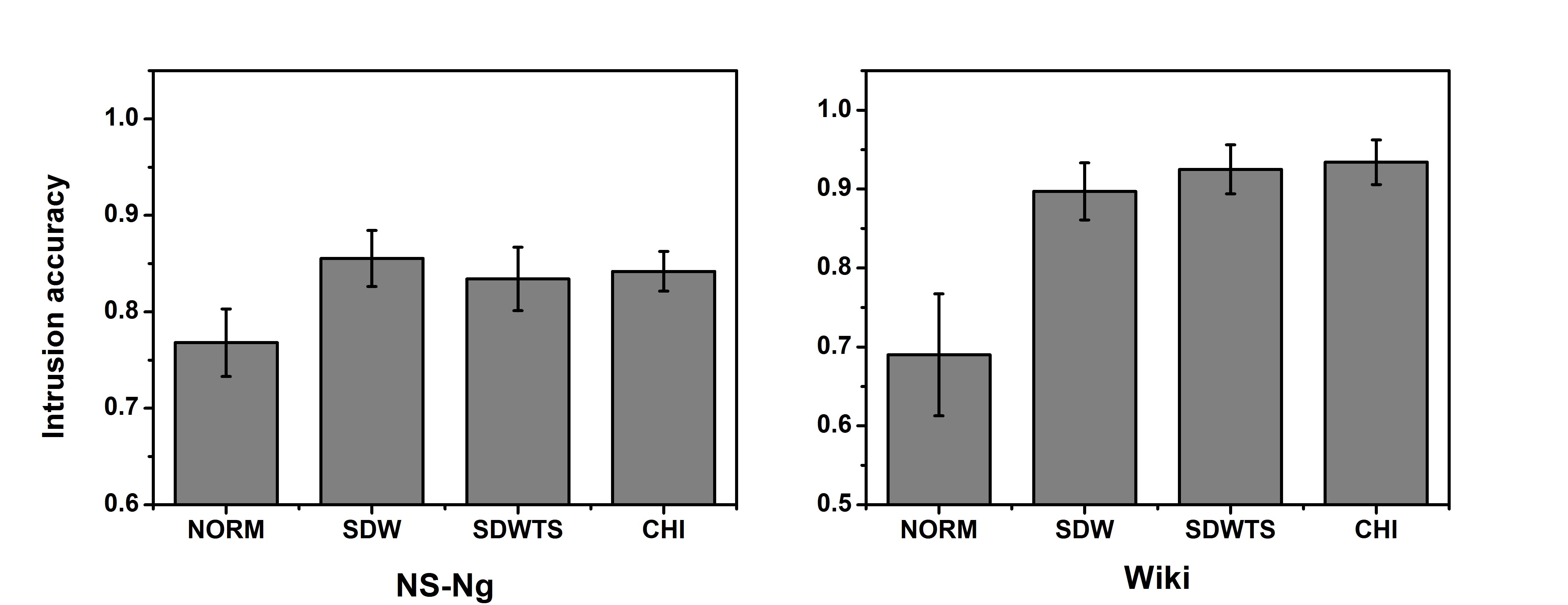}}
\caption{Performance of S\_TOPIC test on NS-Ng (left) and Wiki (right).}
\label{Fig1}
\end{center}
\vskip -0.2in
\end{figure}

\begin{table}[t]
\renewcommand\arraystretch{1.4}
\caption{The S\_TOPIC intrusion task details of two lists of Wiki topics for all four topic representations. The words in bold are the true intruder words; and the words in box are the intruder words selected by the automatical intruder detector \citep{AutoEva2014}.}
\label{sample-table}
\vskip 0.15in
\begin{center}
\begin{small}
\begin{tabular}{p{0.15\columnwidth}|p{0.35\columnwidth}|p{0.38\columnwidth}}
\hline
Representation & Top word list &  Top word list\\
\hline
NORM    & \ovalbox{don} \textbf{rusian} home told father house   &\ovalbox{school} world america students americans \textbf{annotate}\\
SDW     & started \ovalbox{\textbf{rusian}} told wife father bed  &school students schools \ovalbox{\textbf{annotate}} america education   \\
SDWTS   & started home told \ovalbox{\textbf{rusian}} father bed  &school students schools movement \ovalbox{\textbf{annotate}} education   \\
CHI     & started home told \ovalbox{\textbf{rusian}} father bed  &\ovalbox{\textbf{annotate}} schools movement america education   \\
\hline
\end{tabular}
\end{small}
\end{center}
\vskip -0.1in
\end{table}

\paragraph{Performance on S\_SELF test} In this test we select intruder words from sub-top (from 11th to 100th) words in the current topic. Figure 4 shows the average performance (on NS-Ng) of every ten adjacent different sub-top words, which are used as intruder words. Overall, we can see that our reranking representations perform a little better than the NORM representation, e.g., 0.61 (NORM) vs. 0.65 (SDW) on the 91th to 100th sub-top intruder words. This is another evidence that our reranking methodologies can top more representative words.

\begin{figure}[t]
\vskip 0.2in
\begin{center}
\centerline{\includegraphics[width=0.95\columnwidth]{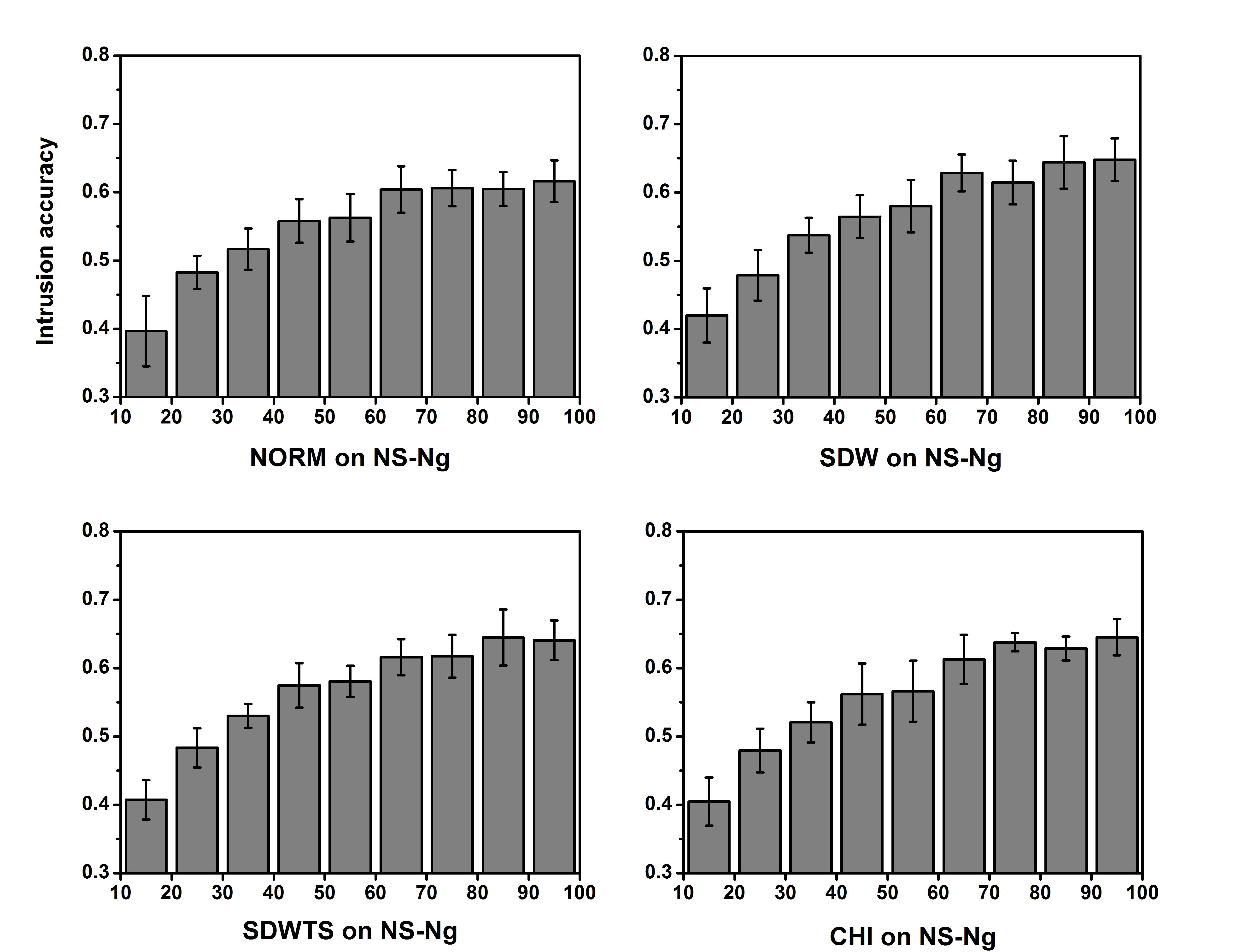}}
\caption{Performance of S\_TOPIC test on NS-Ng.}
\label{Fig1}
\end{center}
\vskip -0.2in
\end{figure}

\section{Related Work}

In topic modeling research, the topic is the word mixture proportion in form. How to express topics is an active direction. Basically, topics are represented as lists of top-\emph{M} words with highest marginal probability. In some early researches, topics are manually labeled for convenient presentation of research results \citep{Mei2005}. Recent attempts on topic representation are focusing on automatical topic labeling algorithms \citep{Labeling2007,Bestword2010,labeling2011,Labeling2013,Al2014,Topic2015}. Most of them are based on the standard top-\emph{M} word list representation, and generate candidate topic labels using external knowledge resources. Some typical works are illustrated as follows: Automatical topic labeling proposed in \citep{labeling2011} generates candidate labels from Wikipedia. It first uses the top-10 topic words to query Wikipedia for relevant article titles, and then uses these titles to generate secondary candidate labels. Finally, the candidate labels are ranked by a supervised model. The authors of \citep{Labeling2013} propose an algorithm to label topics by images \cite{yang2015,yang2016}, instead of text. This algorithm collects candidate image labels by querying Google with top-5 topic words, where the search is restricted to the English Wikipedia. Textual information from the metadata and visual features (e.g., SIFT descriptor) extracted from images \citep{lin2018,yang2017a,yang2017b} are used in ranking the candidate image labels. Besides, we see an interesting comparison among different topic representations within a document retrieval task \citep{Topic2015}.

Nowadays, mainstream topic representations are based on the standard top-\emph{M} word list, and moreover, the automatical topic quality metrics \citep{Intrusion2009,Coh2011,Eva2013} are also based on it. Compared with the standard top-\emph{M} word list, our reranking methodologies further take this factor into consideration. The reranking top-\emph{M} words are, empirically, much more representative for a given topic. The automatical topic labeling in \citep{Bestword2010} also follows a reranking idea. However, it only reranks the topic words in the top-\emph{M} list. In contrast, our methodologies rerank the vocabulary.

\section{Conclusion}

We investigate how to find more representative word lists to represent topics learnt by topic models. We propose three methodologies to rerank topic words by considering the marginal probability on words over different topics. The reranked top-\emph{M} words are used as novel topic representations, namely SDW, SDWTS and CHI representations. Experimental results indicate that our methodologies can (1) effectively filter out stopword-like words and (2) find more representative topic words comparing with the standard topic words with highest marginal probability.

{\small
\bibliographystyle{unsrt} 
\bibliography{ref1}
}

\end{document}